\documentstyle[aps,prl,epsfig,floats,twocolumn]{revtex} 

\begin{document}

\title{ Shot Noise Suppression at 2D Hopping}

\author{Viktor A. Sverdlov, Alexander N. Korotkov and Konstantin K. Likharev }

\address{
Department of Physics and Astronomy, State University of New York,
Stony Brook, New York 11794-3800 }

\date{\today}
\maketitle
\begin{abstract}
We have used Monte Carlo simulation  
to calculate the shot noise intensity $S_I(\omega)$ at 2D hopping 
using two models: a slanted lattice of localized sites with equal
energies and 
a set of localized sites with random positions and energies. 
For wide samples we have found a similar  
dependence of the Fano factor $F \equiv S_I(0)/2eI$ on 
the sample length $L$: $F \propto L^{-\alpha}$  
where $\alpha =0.85 \pm 0.02$ and 
$\alpha = 0.85\pm 0.07$ in uniform and random models, respectively.
Moreover, at least for the uniform model, all the data for $F$ as the function
of sample length $L$ and width $W$ may be presented via 
a single function of the ratio $W/L^{\beta}$, with 
$\beta = 2\alpha-1 \approx 0.7$. This relation has
been interpreted using a simple scaling theory. 
\end{abstract}
 


        Shot noise at electron transport has been the subject 
of intensive experimental and theoretical research lately (for a recent 
review see, e.g., Ref. \cite{Buttiker}), 
because it may provide important information about nonequilibrium properties 
of conductors, unavailable from other transport characteristics. Another 
motivation for  studies of shot noise is its direct relation 
to electric charge discreteness. Namely, the 
smallness of the spectral density of current fluctuations at low frequency, 
$S_I(0)$, in comparison with the
Schottky value of $2eI$, is a necessary condition for quasi-continuous 
charge transfer \cite{Av-Likh,Matsuoka}. Such ``sub-electron" transfer 
through conductors with sufficiently high resistance $R$ and low stray 
capacitance $C$ may make possible several resistively-coupled single-electron 
devices insensitive to background charge randomness \cite{Appl}. 
In this context, hopping conductors are very promising, so that the 
development of understanding of shot noise in such conductors seems 
to be an important task.

        However, though the basic theory of hopping conductivity is well 
developed \cite{book}, until recently little had been known about noise 
at hopping. Few publications we were aware of had been devoted to narrowband, 
$1/f$-type noise (see, e.g., Ref. \cite{Kogan} and references therein) 
rather than broadband fluctuations such as shot noise. This is why in the 
recent work of our group \cite{Sasha} a detailed 
theoretical study of broadband current fluctuations at 1D hopping was  
carried out (on the foundation of prior important work on statistics of the 
so-called Asymmetric Simple Exclusion Process (ASEP) model \cite{Derrida}). 

        For uniform, linear 1D arrays the low-frequency noise 
depends on the boundary conditions (namely, the filling factors 
$f_L, f_R$ of the edge sites) and may or may not be dominated 
by boundary bottlenecks. 
In the former case, the Fano factor $F=S_I(0)/2eI$  tends 
to a finite value of the order of 1 (e.~g., for $f_L = f_R = f$, $T=0$ 
and negligible Coulomb interaction, $F=|1-2f|$) i.e., shot noise  
suppression is insignificant. In the absence of boundary bottlenecks 
(e.~g., if $f_L = f_R = 1/2$), the Fano factor tends to zero 
at large number of hops $N$, but only as $N^{-1/2}$, 
i.e.\ much slower than in 1D arrays of tunnel junctions where $F = 1/N$ 
far enough from the Coulomb blockade threshold \cite{Matsuoka,Sasha}. 
(This behavior has been explained \cite{Sasha} using a simple scaling theory 
which also explains other features, i.e. the frequency dependence 
$S_I(\omega) \propto \omega^{-1/3}$ in an intermediate 
frequency range.)  
Nonuniformity of 1D hopping systems decreases the noise suppression, 
bringing the Fano factor closer to the Schottky value $F = 1$. 

        The goal of this paper is to show that the ability of electrons to
circumvent transport bottlenecks 
at 2D hopping leads to a qualitatively 
different situation. Namely, in sufficiently long and broad samples  
the shot noise may be suppressed quite considerably: 
\begin{equation}
F \propto L^{-\alpha},
\label{power}
\end{equation}
where $L$ is the conductor length and $\alpha \approx 0.85$, even in 
ultimately nonuniform conductors. 

        We have employed the usual Monte-Carlo simulation 
technique (see, e.g., 
Ref.\ \cite{Sasha}) to analyze two different models, so far both without 
Coulomb interaction and at vanishing temperature: 

         - Model A: hopping between sites with random localization energies, 
randomly distributed over a 2D sample, and

        - Model B: hopping on a uniform, slanted lattice without site energy
fluctuations. 

        In both models, each site may be occupied by just one electron, and 
the rate of (inelastic) transitions between the sites is described by the 
usual formula (see, e.g., Eq. (4.2.17) in Ref. \cite{book}):
\begin{equation}
\Gamma_{i\rightarrow j}=A \, \frac{\epsilon_{ij}}
{1-\exp(-\epsilon_{ij}/T)},
\label{gamma}
\end{equation}
corresponding to the constant density of phonon states. Here $\epsilon_{ij}$ 
is the electron energy gain during the hop $i\rightarrow j$; in the absence 
of Coulomb interaction between electrons this gain can be expressed as 
	\begin{equation}
\epsilon_{ij}=(\epsilon_i-\epsilon_j)-eE(x_i-x_j),
	\end{equation}
where $E$ is an external electric field applied along the $x$ axis.  For 
relatively short samples special care should be taken to adequately describe 
electron transfer between the electrodes and the edge 
localized sites. After experimenting with various options, we have 
concluded that the same expression (\ref{gamma}) may be used to describe 
this transfer, without creating unphysical bottlenecks at the 
electrode-sample interfaces \cite{footnote}. 

        In our main Model A, single-particle site energies $\epsilon_i$ are 
distributed randomly within a broad energy band, with a constant 2D 
density of states $D$, and site positions ${x_i,y_i}$ are randomly 
distributed within 
a rectangular sample of length $L$ and width $W$. The rate amplitude $A$ 
is an exponential function of the intersite distance $r_{ij}$:
\begin{equation}
A = A_0 \exp(-r_{ij}/a),  
\label{A}
\end{equation}
where $a$ is half of the localization radius. All the results 
have been averaged not only over a sufficiently long time period, 
but also over a set of random 
samples with the same global dimensionless parameters $L/a$, $W/a$, 
and $eEDa^3$ (parameter $A_0$ just determines the scale 
$I_0 = eA_0W/Da^3$ of the total current).
	Such averaging requires considerable computer resources; 
the calculations have been performed on IBM SP parallel supercomputer.

\begin{figure}
\begin{center}
\vskip-0.3cm
\epsfig{figure=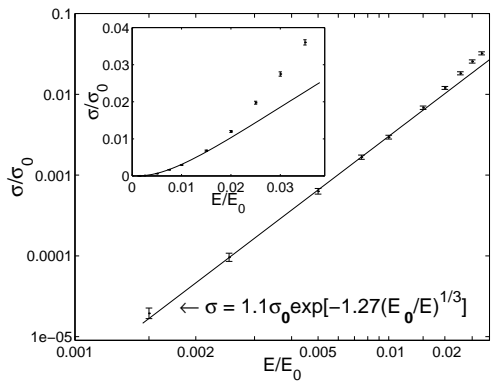,scale=0.35} 
\vskip-0.2cm
\end{center}
\caption{The nonlinear dc conductivity $\sigma =I/WE$ (normalized by 
        $\sigma_0 = I_0/WE_0$), as a function of electric field $E$ for 
        Model A. The straight line is the analytical fit discussed in 
        the text. Inset: the same data on a linear scale.} 
\label{fig1}\end{figure}        

        Figure 1 shows the numerically calculated nonlinear dc conductivity  
$\sigma =I/WE$ as a function of electric field 
$E$ for sufficiently long and wide samples. Depending on $E$, we have 
simulated samples of area $L\times W$ ranging from $120 a 
\times 30a$ up to $300 a\times 120 a$ to keep the number of ``active'' sites 
$N_s \sim 1500$ approximately constant (the growing error bars at lower $E$ 
are due to larger fluctuations from sample to sample). The absence 
of significant dependence of $\sigma$ on the sample size was being checked. 
For low electric fields $E \lesssim 0.02 E_0$ ($E_0 = 1/eDa^3$) the current 
follows 
closely the dependence $I/E \propto \exp[- C (E_0/E)^{1/3}]$ 
expected for 2D variable range hopping \cite{book,Pollak} in the 
activationless (``high field'') regime. 
The best fit (straight line in Fig.\ 1) gives the numerical constant 
$C\approx 1.27$. (This number is to be compared with the value
$C =1.02$ following from analytical calculations in Ref.\ \cite{Thompson}.)  
The minor deviation from the analytical dependence  
at $E \gtrsim 0.02 E_0$ is possibly due to 
multiple, well-branched percolation paths which are not considered 
in the usual theoretical treatment of high field hopping. 

\begin{figure}
\begin{center}
\epsfxsize=2.5in
\vskip-0.3cm
\epsfig{figure=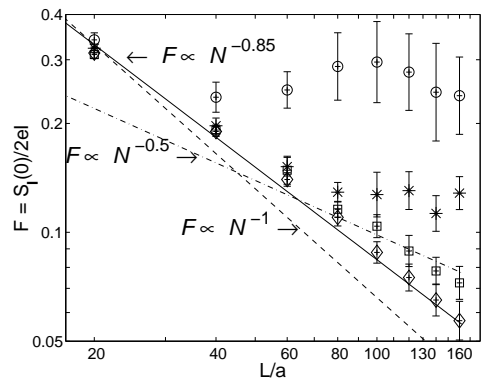,scale=0.35} 
\vskip-0.1cm
\end{center}
\caption{Dependence of the Fano factor $F$ in Model A
        (averaged over 32 sample realizations) on the sample length 
        $L$ for $E=0.035E_0$. Circles: $W=5a$, asterisks: $W=10a$, 
        squares: $W=20a$, diamonds: $W \rightarrow \infty$. Error 
        bars show the standard deviation of the mean.}
\label{fig2}\end{figure}

Figure 2 shows the Fano factor (averaged over 32 sample realizations) 
as a function of the sample length for $E=0.035E_0$ and for tree values 
of sample width: $W=5 a$ (circles), $W=10 a$ (asterisks) and $W=20 a$ 
(squares). 
At this field the average hop length is $\overline{r}\approx 3.3 a$
(with r.m.s.\ projections $r_x\approx 3.1 a$, 
$r_y\approx 1.9 a$) so that 
the factor $\exp (-\overline{r}/a)$ is still small.
Most importantly, we see that shot noise is suppressed considerably 
($F \ll 1$) in  wide and long samples. 
As a function of the sample length $L$ at fixed width $W$, the average 
Fano factor first decreases following Eq.\ (\ref{power}) (solid line
in Fig.\ 2) and then at certain length deviates up from this 
dependence. The deviation starts at larger $L$ for wider samples. 
For narrower samples one can see the saturation of the average 
Fano factor at large $L$; simultaneously the width of the Fano factor 
distribution grows significantly. 
The average Fano factor also decreases with the sample width at fixed length, 
obviously saturating at large $W$ since the electron transport in remote 
parts of a very wide sample is uncorrelated. 
Having performed the calculations at fixed length for several widths (not
all the results are shown in Fig.\ 2) we have extrapolated the Fano factor 
dependence to infinitely wide 2D samples.   
As a function of $L$, these results (shown in Fig.~2 by diamonds) 
closely follow Eq.\ (\ref{power}) with $\alpha = 0.85 \pm 0.07$. 
The power-law dependence $F(L)$ has been observed within one order of 
magnitude range of $L$. The longer samples have not been studied due
to computer limitations (as an example, calculations of a single point 
$L=160 a$, $W=20 a$ in Fig.\ 2 has required 380 hours of total CPU time).

        In order to verify the shot noise suppression for larger
set of lengths and widths, we have used 
the simplified Model B in which $(N-1)\times M$ localized sites with 
equal energies are arranged on a uniform slanted square lattice 
(see inset in Fig.\ 3) \cite{rect}. In accordance with Eq.\ (\ref{gamma}), 
at $T=0$ the transport is 
unidirectional, and transfer rates $\Gamma$ between all the internal 
neighboring sites are equal. For the links from the left electrode
to the nearest internal sites we have selected the rates $2\Gamma f_L$
(the factor of 2 reflects two 
``channels'' per internal site) while the rates of hopping onto 
the right electrode are $2\Gamma (1-f_R)$. For the numerical analysis we
have chosen the case $f_L=f_R=0.5$ in which, similarly to 1D model, 
there are no boundary bottlenecks.
Since Model B does not require averaging over different random realizations, 
it may be studied with much better accuracy using the same computer 
resources. 

\begin{figure}
\begin{center}
\vskip-0.2cm
\epsfig{figure=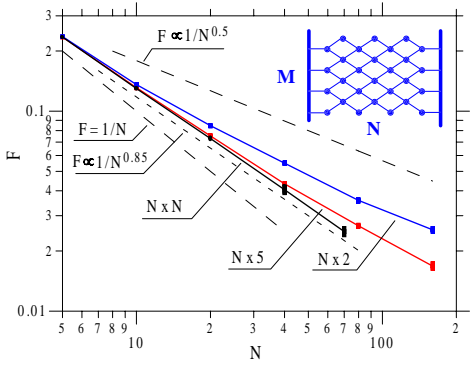,scale=0.43} 
\vskip-0.1cm 
\end{center}
\caption{Fano factor dependence on the length $N$ for 
        hopping on a uniform slanted square lattice (Model B, inset) with 
        width $M$ equal to 2, 5, and $N$.} 
\label{fig3}\end{figure}        

        Figure 3 shows the Fano factor as a function of the array 
length $N$ for several values of width $M$. Again, we see a strong shot 
noise suppression. For sufficiently wide samples 
the suppression follows Eq.\ (\ref{power}) (where now $L$ should be 
replaced by $N$), with similar exponent as in the wide random samples: 
$\alpha = 0.85 \pm 0.02$. 
On the other hand, for fixed $M$ and sufficiently long $N$ 
the suppression power approaches $0.5$, i.e.\ the same value as for 1D 
hopping \cite{Sasha}. 

        There is numerical evidence (see Fig.\ 4), as well as 
 scaling arguments (see below), that for 
$N\gg 1$ and $M\gg 1$ the crossover between these 
two asymptotic laws may be parameterized in the following way: 
        \begin{equation}
F(N,M)=N^{-\frac{1}{2}} M^{-\frac{1}{2}}g(N^{\beta}/M),
        \label{scaling} 
\end{equation}
where $\beta=0.70\pm 0.04$, and the function $g(x)$ 
is shown in Fig.\ 4: 
\begin{equation}
g(x) \propto \left\{\begin{array}{ll}
const,& x\gg1\\
x^{-1/2},&x\ll1.
\label{g}
\end{array}\right.
\end{equation}

        Checking if the data from Fig.\ 2 for the random Model A may 
also be collapsed on the similar universal 
curve we have found a reasonable fit for relatively wide samples
(see inset in Fig.\ 4)  
for the following replacements: $N = L/5.9a$, $M = W/11a$
(for the particular field $E=0.035E_0$). 

        We will start the interpretation of our findings from the uniform 
Model B. Similarly to the 1D ASEP model \cite{Derrida}, in the absence 
of lateral boundary effects due to finite $M$, and at $f_L=f_R=f$ 
the probability of any charge configuration is expected to be the same 
as if each site had independent occupation with probability $\rho =f$
\cite{periodic}. 
As a result, 
dc current between any two neighboring sites should equal 
$\Gamma \rho (1-\rho )$, so the total dc current is
\begin{equation}
I= 2M \Gamma \rho (1-\rho ).
\label{current}
\end{equation} 

\begin{figure}
\begin{center}
\epsfxsize=2.5in
\vskip-0.3cm
\epsfig{figure=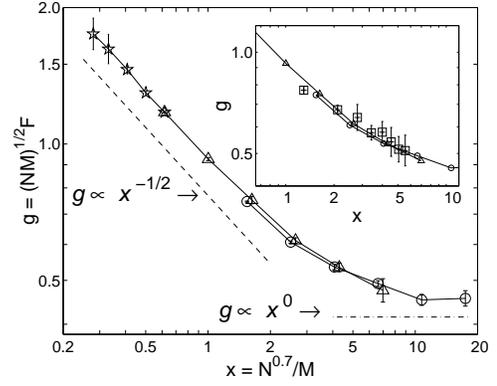,scale=0.35} 
\vskip-0.1cm
\end{center}
\caption{The data from Fig.\ 3 (circles: $M=2$, triangles: $M=5$,
        and stars: $M=N$) collapsed onto universal curve $g(x)$ using 
        parameterization $x\equiv N^{0.7}/M$ and $g\equiv (NM)^{1/2}F$. 
        Lines connecting points are just guides for the eye. 
        The inset shows the same data with added points for $W=20a$ (squares) 
        from Fig.\ 2 using the rescaling $N=L/5.9a$
        and $M=W/11a$. } 
\label{fig4}\end{figure}        

        Following the arguments of Ref.\ \cite{Sasha} we obtain 
only a minor suppression of low-frequency shot noise, 
$F\approx |1-2\rho |$, in the case  
$\rho \neq 0.5$ (the noise significantly decreases at frequencies 
$\omega \gtrsim \omega_l \sim \Gamma |1-2\rho |/N$). 
However, when the coupling with electrodes is strong 
enough, $f_L\geq 0.5$, $f_R\leq 0.5$, the half-filling is 
expected inside the array, $\rho =0.5$, and the Fano factor can indefinitely 
decrease with the array length $N$. 
        In this case, for sufficiently large $N$ (narrow array) 
we may repeat all 1D scaling arguments of Ref.\ \cite{Sasha} based on 
Eq.\ (\ref{current}), and arrive at the following estimates: 
        \begin{equation}  
F \sim (NM)^{-1/2}, \,\,\,  \omega_l \sim \Gamma N^{-3/2}M^{-1/2},  
        \end{equation} 
for the Fano factor and the saturation frequency $\omega_l$, 
above which the dependence $S_I(\omega )/2eI \sim  
(\omega/\Gamma )^{-1/3}N^{-1}$ is expected. 

        Obviously, this result should eventually fail if we start
increasing the width $M$ for fixed $N$, because $F$ cannot depend on $M$ 
in the limit of wide array (since the transport in remote parts of the array 
is uncorrelated and so both $S_I$ and $I$ are additive). 
Denoting the crossover width as $M_0$ we get the
estimate $F\sim (NM_0)^{-1/2}$ for wide arrays. 
It would be natural to expect $M_0\propto N$, however, the numerical results
(Fig.\ 3) indicate the power-law dependence, $M_0\propto N^\beta$, with 
$\beta$ being a phenomenological parameter. If this assumption is true, 
we obtain the following estimates: 
        \begin{equation}
F \sim N^{-(1+\beta )/2}, \,\,\,  \omega_l \sim \Gamma N^{-(3+\beta )/2},  
        \label{Fwide}\end{equation} 
for wide arrays, $M\gg N^\beta$. 
Our numerical result, $\alpha =0.85 \pm 0.02$, then leads to the value 
$\beta =2\alpha - 1 = 0.7 \pm 0.04$. 
        For intermediate widths it is natural to suggest that the crossover 
is governed by some function of the ratio $x \equiv N^{\beta}/M$ alone.
Thus we recover the behavior described by Eqs.\ 
(\ref{scaling})--(\ref{g}) and illustrated by Fig.\ 4. [Actually, 
at this stage we cannot rule out the possibility that the function 
$g$ in Eq.\ (\ref{scaling}) also has a weak dependence on $M$ that
would lead to either smoothing or sharpening of the curve in 
Fig.\ 4, leaving however the asymptotes (\ref{g}) intact.] 

        If we apply the same scaling arguments which have led to 
Eq.\ (\ref{Fwide}), to smaller blocks 
\cite{Sasha} of size $N_\omega \times N_\omega^\beta$, we obtain 
the noise frequency dependence
        \begin{equation}
S_I(\omega ) /2eI \sim \omega^{-\gamma} N^{-1}, \,\,\,
\gamma \equiv  (1-\beta )/(3+\beta ) , 
        \end{equation}  
for wide arrays at intermediate frequencies, $\omega_l \ll \omega \ll \Gamma$.

This power-law dependence, $S_I(\omega )\propto \omega^{-0.08}$,   
was confirmed numerically with the accuracy of 
the exponential factor about $\pm 0.007$. The same frequency dependence 
should be expected for the narrow arrays, $M\ll N^\beta$, at frequencies 
higher than $\sim\Gamma M^{-(3+\beta )/2\beta} $, while at lower frequencies 
$S_I(\omega )\propto \omega^{-1/3}$, as discussed above.

        Now, let us turn back to our main, random Model A. 
According to the percolation picture of hopping \cite{book}, the conductivity
of a sample is determined by a ``percolation cluster'', essentially 
a network of sites connected by the most probable hop paths. The percolation 
cluster may be divided into blocks of a certain size, such that the aggregate 
characteristics of each block (e.g. the average current) are nearly equal, 
even though inside each block the sample is highly (exponentially) 
nonuniform. Average block size in the transport direction for the 
random model A may be determined by mapping 2D hopping in disordered wide
samples onto the uniform model B, for $M>N$, where the shot noise suppression
was found  to be similar. If this interpretation 
is valid, then the ratio $L/N=5.9 a$ obtained from the mapping should be 
comparable to the correlation length 
$L_c \sim \overline{r} (\overline{r}/a)^\nu \sim 3.3 a$  \cite{Shklovskii76}. 
We believe these numbers are reasonably consistent.

On the other hand, the model A behavior in narrow samples ($ W \lesssim 10 a$)
is quite different from that in the uniform model: instead of going down with
growing $N$, the Fano factor saturates. Simultaneously, the statistics of
$F$ becomes much wider. This behavior is very natural, since if the sample 
is narrower than the block size (for our value of electric field, about 
$10 a$), the exponentially broad distribution of hopping paths within the 
block is revealed and is mapped onto the properties of the sample as a whole. 

        Our results for wide samples are in good agreement with 
data from a recent experiment 
\cite{Volodya} in which shot noise at hopping was measured 
in $p-$type SiGe quantum wells. Actually, the experimental {\it I-V} curve 
significantly differs from that in our Model A. 
However, our result (\ref{power}) for $F$ seem more general. 
For wide samples with two different lengths $L_1=2\mu$m and $L_2=5 \mu$m 
the Fano factor was measured \cite{Volodya} to equal 
$F_1=0.43$ and $F_2=0.2$, respectively. This corresponds  
to Eq.\ (\ref{power}) with $\alpha =0.84$, the value which is virtually 
equal to our result $\alpha =0.85$. Such perfect agreement is possibly 
just a coincidence, since so far only two experimental points are 
available. Evidently, it would be valuable to have more experimental data, 
in order to verify 
the shot noise suppression power $\alpha =0.85$. 

        In conclusion, we have numerically investigated shot noise 
suppression at 2D hopping in random samples as well as in uniform arrays.
Very similar shot noise suppression has been found for both models for wide
and long samples. At 2D hopping the Fano factor decreases with 
the sample length as 
$F \propto 1/L^{0.85}$. 
This suggests 
that shot noise 
suppression at 2D hopping is insensitive to details of the hopping process, 
e.g., the energy dependence of the hopping rate. If this surmise is true, 
the Fano factor should not be a very strong function of temperature. 
It may, however, be substantially altered by Coulomb interaction 
of hopping electrons, as in the 1D case \cite{Sasha}. Our next plans 
are to explore the effects of both these factors.

        Fruitful discussions with V.~Kuznetsov and generous help by J.~Wells
are gratefully acknowledged. 
The work was supported in part by the Engineering Research Program of the
Office of Basic Energy Sciences at the Department of Energy. The authors also
acknowledge the use of the Oak Ridge National Laboratory IBM SP computer, 
funded by the Department of Energy's Offices of Science and Energy Efficiency
program.


\begin{thebibliography}{99}


\bibitem{Buttiker} Ya. M. Blanter and M. B\"uttiker,
        Physics Reports {\bf 336}, 2 (2000).

\bibitem{Av-Likh} D. V. Averin and K. K. Likharev, in: {\it Mesoscopic
        phenomena in solids}, ed. by B. Altshuler et al. (Elsevier, 
        Amsterdam, 1991), Ch. 6.

\bibitem{Matsuoka} K. A. Matsuoka and K. K. Likharev, Phys. Rev. B {\bf 57},
        15613 (1998).

\bibitem{Appl} K. K. Likharev, Proc. IEEE {\bf 87}, 606 (1999); 
        A. N. Korotkov, Int. J. Electron {\bf 86}, 511 (1999). 

\bibitem{book} B. I. Shklovskii and A. L. Efros, {\it Electronic
        properties of doped semiconductors} (Springer, Berlin, 1984).

\bibitem{Kogan} Sh. Kogan, Phys. Rev. B {\bf 57}, 9736 (1998).

\bibitem{Sasha} A. N. Korotkov and K. K. Likharev, Phys. Rev. B {\bf 61},
	15975 (2000). 

\bibitem{Derrida} B. Derrida and E. Domany, J. Stat. Phys. {\bf 69}, 
        667 (1992).

\bibitem{footnote} Physically, this transfer may be dominated by elastic 
        tunneling for which Eq.\ (\ref{gamma}) is not quite adequate. 
        Our study, however, was focused on long samples ($L\gg a$) 
        for which this difference is unimportant.

\bibitem{Pollak} M. Pollak and I. Riess, J. Phys. C: Solid State Phys.
        {\bf 9}, 2339 (1976.)

\bibitem{Thompson} R. B. Thompson and M. Singh, Phil. Mag. B {\bf 75}, 
        293 (1997). 

\bibitem{rect} The uniform model on the usual (non-slanted) 2D array 
        cannot give an adequate presentation of hopping in random systems, 
        since it fails to describe any hops other than exactly along 
        the applied field, and hence may be exactly reduced to the 
        uniform 1D model studied in Ref.\ \cite{Sasha}.

\bibitem{periodic} This result becomes exact even for finite $M$ at 
        the cyclic boundary conditions in the direction normal to the field.  

\bibitem{Shklovskii76} B. I. Shklovskii, Fiz. Tekh. Poluprovodn. {\bf 10},
        1440 (1976) [Sov. Phys. Semicond. {\bf 10}, 855 (1976)].

\bibitem{Volodya} V. V. Kuznetsov, E. E. Mendez, E. T. Croke,   
X. Zuo, and G. L. Snider, 
        Phys. Rev. Lett. {\bf 85}, 397 (2000)


\end{thebibliography}
\end{document}